\documentclass[useAMS,usenatbib]{mn2e}
\usepackage{graphicx}
\usepackage[letterpaper]{geometry}
\usepackage{bm}

\author{Steve Price, Bennett Link, Richard I. Epstein, Hui Li}
\title{Thermo-Resistive Instability in Magnetar Crusts}
\date{\today}

\begin{document}

\onecolumn

\maketitle

\begin{abstract}
We investigate a thermo-resistive instability in the outer crusts of magnetars wherein a perturbation in temperature increases ohmic heating.  We show that magnetars of characteristic age $\tau_{\rm{age}} \sim 10^4 $ yr are unstable over timescales as short as days if strong current sheets are present in the outer crust.  This instability could play an important role in the thermal and magnetic field evolution of magnetars, and may be related to bursting activity in magnetars.
\end{abstract}

\section{Introduction}

Soft gamma repeaters (SGRs) are a class of highly magnetized neutron
stars (magnetars) that exhibit persistent X-ray emission,
interrupted periodically by short bursts of gamma rays.  The more frequently occurring short
bursts have typical durations of 0.01 - 1s, with peak luminosities of $\sim
10^{42} \, \rm{erg} \,\rm{s}^{-1}$.  Giant flares are much more rare and energetic outbursts, with peak luminosites $10^2 - 10^{5}$ times larger than the
short bursts.  Three giant flares have been observed to date: from SGR
0526-66 on March 5, 1979 (\cite{helfand79}; \cite{mazets79}), SGR
1900+14 on August 27, 1998 \citep{hurley99}, and SGR 1806-20 on
December 27, 2004 \citep{hurley05}.  SGR 1806-20 has produced the largest observed giant flare, releasing $\sim 10^{46} $ erg of energy over 380 s.  Precursors to
two of the three giant flares have been identified.  The 2004 giant flare was
preceded by a 1s long event, 140 s prior to the initial hard spike
\citep{hurley05}.  A similar precursor to the August 27 giant flare
was observed \citep{ibrahim01}, with duration $\sim .05 \,$s preceding
the initial spike by 0.4 s.

SGR flares are thought to represent the release of magnetic energy, though the trigger mechanism remains uncertain.  \cite{td95} propose that the short bursts
observed from SGRs are the result of fracturing the crust of the neutron star by magnetic stresses.  Recent calculations indicate that neutron
star crusts fail catastrophically under stress \citep{horowitz}.  The rigid crust could act as a gate, releasing magnetic energy when magnetic stresses cause the crust to fail.  Larger events such as giant flares could be caused by large scale readjustment of the magnetic field due to magnetic instability \citep{td95}.  As the field readjusts, magnetic reconnection leads to the formation of a pair plasma which is injected into the magnetosphere \citep{td95}.  Alternatively, \cite{lyutikov} argued that the short rise time of the 2004 giant flare requires that the magnetic energy be stored in the magnetosphere rather than the stellar interior.  In this scenario, slow untwisting of the internal magnetic field eventually leads to sudden relaxation of the twist in the magnetospheric field, releasing the energy necessary to power the flare.

The initial configuration and subsequent evolution of magnetic fields in highly-magnetized neutron stars is a complicated problem.  The field evolves continuously due to the effects of ohmic decay, ambipolar diffusion, and Hall drift. Recently, \citet{ponsgeppert} studied the evolution of magnetic fields in neutron star crusts, emphasizing the importance of Hall drift.  Their results indicate that Hall drift of crustal fields can create small-scale magnetic field structures, and that those structures can drift to regions of higher resistivity. The simulations of \cite{ponsgeppert} were restricted to magnetic fields in the inner crust.  

In this paper, we focus on the outer crust, and show that large currents can lead to a thermo-resistive instability, affecting the thermal evolution of the star.  As the instability evolves, large portions of the crust may melt, allowing the magnetic field to evolve on a timescale much faster than the average ohmic decay and Hall timescales.  The enhanced magnetic evolution resulting from instability may be related to flare activity in magnetars.

This paper is organized as follows.  In section 2, we describe the relevant physics of the thermo-resistive instability in neutron star crusts.  In section 3, we describe the physical processes that lead to evolution of the neutron star magnetic field.  In section 4, we describe our neutron star model and give details of calculations of the instability growth rate.  Section 5 contains discussion and our conclusions.

\section {Thermo-resistive instability}

The dipole fields of magnetars are inferred to be in the range $\sim (0.5-20) \times 10^{14} \,$ G, based on spindown measurements of SGRs and AXPs ( see \cite{mereghetti} for a review).  In order for the magnetic field to be stable in neutron stars, it must contain both toroidal and poloidal components \citep{tayler}.  The toroidal component in a stable configuration has a twisted torus shape, and may be an order of magnitude larger than the poloidal component \citep{braithwaite09}. 

Large crustal currents associated with the toroidal field would
produce significant ohmic heating due to the relatively high
resistivity in the outer crust.  Ohmic heating can account for the
observed trend of surface temperature that increases with surface
field observed in neutron stars with $B
> 10^{13} $ G \citep{ponslink}.  Cooling simulations
show that a heating layer in the outer crust, as would arise from
current decay, can explain the high surface temperatures of
magnetars \citep{kaminker}.  

Current decay is determined primarily by
electron-phonon interactions for a temperature $T$ below the melting
temperature $T_{\rm melt}$. In this regime, the electrical resistivity scales as $T$,
so that a small increase in temperature leads to increased heat
dissipation (Fig. 1).  The additional heating raises the temperature
further, and a temperature runaway may develop if thermal transport is
unable to quench the feedback process.  As we show in this paper,
this instability can occur in the outer crusts of neutron stars,
where the electrical resistivity is relatively high and the thermal
conductivity is low.

\begin{figure}
     \includegraphics[scale=.50]{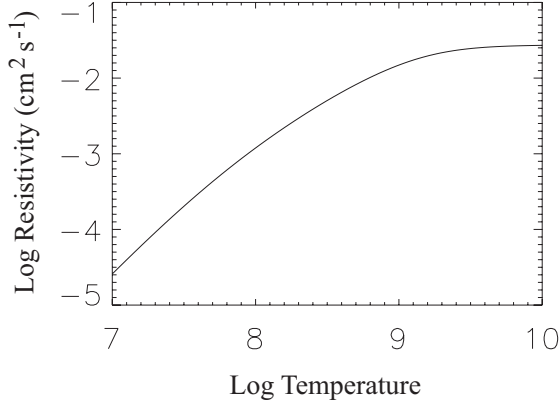}
     \caption{Resistivity of the crust at a density of $10^9 \,\rm{g} \,\rm{cm}^{-3}$.  The melting temperature at this density is $\sim 3 \times 10^8 \,K$.}
\end{figure}

\section{Ohmic decay and Hall Drift}

The evolution of the magnetic field in neutron stars after birth is
determined primarily by ohmic decay and Hall drift.  The ohmic decay timescale is $\tau_{\rm ohm} =\eta^{-1} L^2$, where $L$ is the typical magnetic field length scale
and $\eta$ is the electrical resistivity. A typical value for the outer
crust at temperature $T=10^8$ K is 
\begin{equation}
  \tau_{ohm} = 3 \times 10^5 \, \left (  \frac{L}{1 \, \rm{km}}\right )^2 \left ( \frac{\eta}{10^{-3} \, \rm{cm}^2 \, \rm{s}^{-1}}\right )^{-1} \rm{yr}.
\end{equation}
For a magnetar of age $10^4$ yr, crustal currents from the initial
field should still be present. [We note that the $\tau_{\rm ohm}$ was
smaller early in the star's life, since the temperature and
resisitivity were higher.] 

Hall drift creates small scale magnetic structures in the crust over the Hall timescale
\citep{ponsgeppert}, given by
\begin{equation} \label{eq:halltau}
  \tau_{Hall} = \frac{4 \pi n_e e L^2}{cB}, 
\end{equation}
where $n_e$ is the electron density and $B$ is the magnetic field
strength.  A typical value for the outer crust is
\begin{equation}
\tau_{Hall} \sim  6 \left ( \frac{L}{1 \,\rm{km}} \right )^2 \left ( \frac{B}{10^{15} \,G}\right )^{-1} \,\rm{yr}, 
\end{equation}
Hall drift can concentrate currents in the crust.  The induction
equation, neglecting ohmic dissipation,  is
\begin{equation}
   \frac{\partial \bm{ B}}{\partial t} = - \bm{ \nabla} \times \left ( \frac{c}{4 \pi n_e e} (\bm{ \nabla} \times \bm{ B}) \times \bm{ B} \right ).
\end{equation}
To illustrate how Hall drift may affect the magnetic field, consider a magnetic field in cylindrical
coordinates, with only an azimuthal component which depends on $r$,

\begin{equation}
    \bm{B} = B_{\phi} (r) \bm{\hat \phi}.
\end{equation}
Eq. (4) becomes
\begin{equation}
\frac{\partial \bm{ B}}{\partial t} = 
\bm{ \nabla} \times \left ( \frac{c}{4 \pi n_e e} \frac{B_{\phi}}{r} \frac{\partial (r B_{\phi})}{\partial r} \bm{\hat r} \right ).
\end{equation}
Since the magnetic field has only $r$ dependence, and the quantity
inside the parenthesis is in the $\hat r$ direction, the curl is zero,
and the field is stationary.

Outward drift of the field can replenish currents in the outer crust that are decaying through ohmic diffusion.  In order to get an outward drift of the field, the field must have
z-dependence, as considered by \citep{ponsgeppert}.  This would
correspond to a field strength that varies from the magnetic pole to
the magnetic equator. The induction equation for the
$\phi$-component of the field gives
\begin{equation}
\frac{\partial B_{\phi}}{\partial t} = \frac{2}{\tau_{\rm{Hall}}} \frac{L^2}{r} \frac{\partial B_{\phi}}{\partial z}.
\end{equation}
If $\partial B/\partial z$ is positive, the field in the crust will
increase over a typical timescale of $\tau_{Hall}$. 

We assume henceforth that strong currents exist in the crust at an age
of $\sim 10^4$ years, and explore the consequences.

\section{Calculations}

\subsection{Equations and Boundary Conditions}

We model the outer crust as an infinite slab, with $\bm{\hat x}$ pointing into the star (see Fig. 2).  The thermal evolution of the neutron star outer crust is described by the energy conservation equation,
\begin{equation} \label{eq:energy}
   c_v (\rho,T) \frac{dT (x)}{dt} = \frac{4 \pi \eta (\rho,T)}{c^2}j^2 + \bm{ \nabla} \cdot \left [\kappa(\rho,T) \, \bm{ \nabla} T(x) \right ]   + Q_{\nu} (\rho,T),
\end{equation}
where  $\rho$ is the density, $c_v$ is the specific heat, $\bm{j}$ is the electric current, $Q_{\nu}$ is the neutrino emissivity, and  $\kappa$ is the thermal conductivity.  The magnetic field evolution is described by the induction equation
\begin{equation}
   \frac{\partial \bm{ B}(x)}{\partial t} = - \bm{ \nabla} \times \left ( \eta(\rho,T) \bm{ \nabla} \times \bm{ B}(x) \right ),
\end{equation}
where the magnetic field is related to the current by 
\begin{equation} \label{eq:bfield}
   \bm{ j}(x) = \frac{c}{4 \pi} \bm{ \nabla} \times \bm{ B}(x).
\end{equation}

As justified below, we work in an approximation in which magnetic induction can be ignored, so we need not specify boundary conditions on $\bm{B}$.  The slab consists of of 3 regions - an atmosphere with no magnetic dissipation, an outer crust, and an isothermal inner crust/core.  The atmosphere extends from a density of $10^4 \rm{g} \, \rm{cm}^{-3}$ at the stellar surface to  $10^6 \rm{g} \, \rm{cm}^{-3}$, the outer boundary of the crust.  For the atmosphere and outer crust zones, we employ the density model of \cite{friedman}.  The inner crust/core zone is assumed to be an infinite heat reservoir, beginning at a density of $\rho=10^{11} \rm{g} \, \rm{cm}^{-3}$.

\begin{figure}
     \includegraphics[scale=.80]{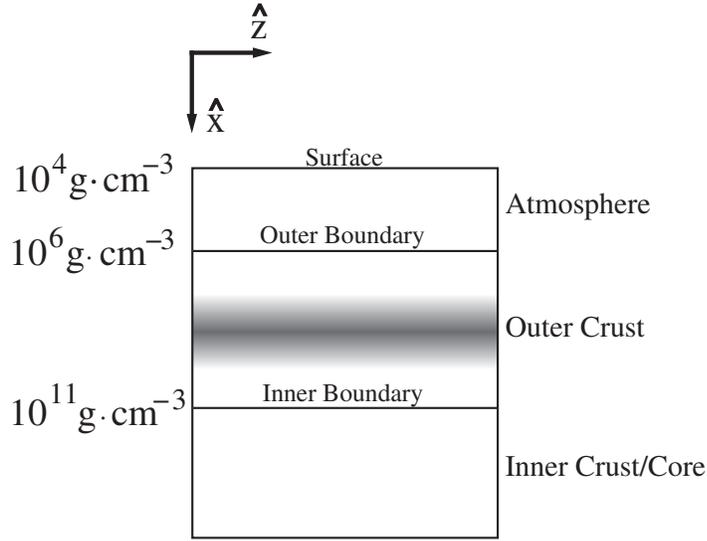}
     \caption{Neutron Star model.  The shading represents the region of ohmic heating.}
\end{figure}

\emph{Boundary conditions.}  At the stellar surface we choose the unperturbed temperature to be,
\begin{equation}
T=T_s,
\end{equation}
and require that the heat flux at the boundary equal the blackbody emission rate at temperature $T_s$,
\begin{equation}
   \kappa \frac{dT_s}{dz} = \sigma T_s^4.
\end{equation}

\begin{figure}
     \includegraphics[scale=0.5]{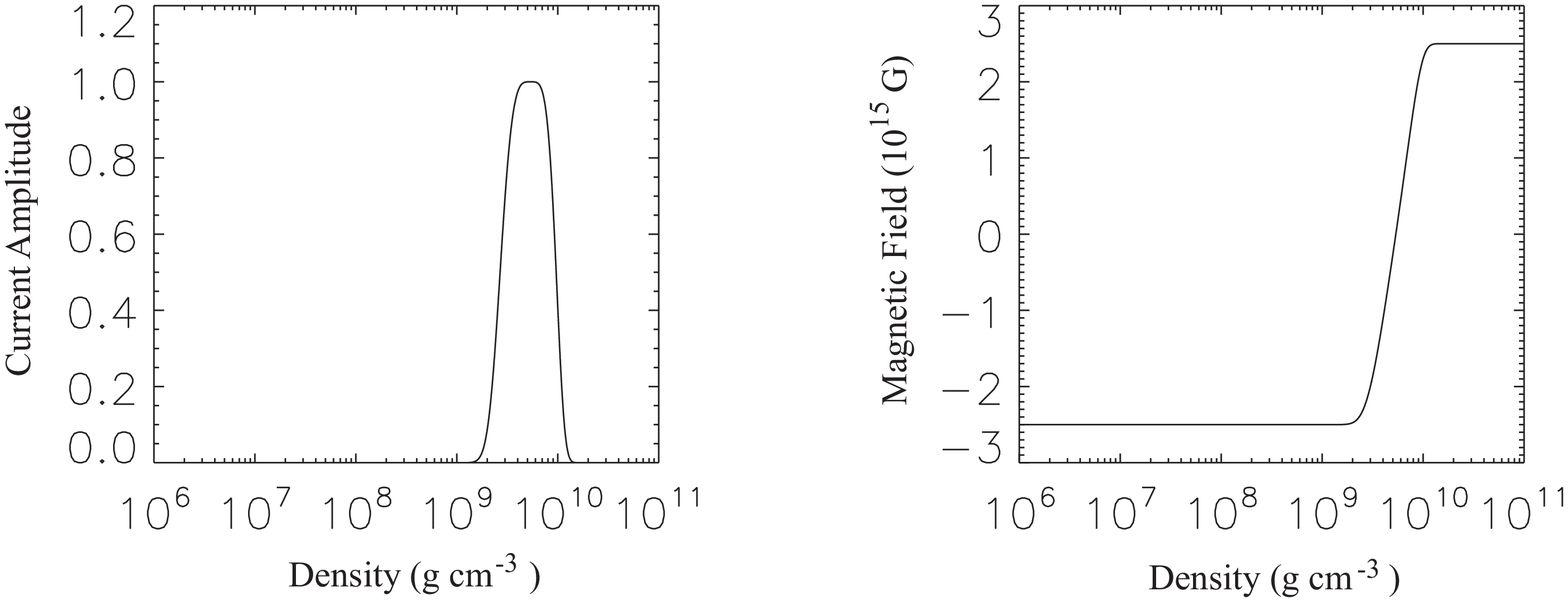} 
     \caption{Sample current sheet and associated magnetic field.  The current peaks at $x_0=100 \, \rm{m}$,approximately at the center of the outer crust.}
\label{fig:current}
\end{figure}

As a simple model of the toroidal component of the neutron star magnetic field, we introduce to the outer crust zone a current sheet of width $L$:
\begin{equation}
   \bm{ j}(x) = j_0\, e^{-(x-x_0/2 L)^4} \bm{\hat z} ,
\end{equation}
where $j_0$ is the amplitude of the current, and $x_0$ the location of the
peak current.  This analytic form allows a large, nearly uniform
current near the heating peak, falling off sharply for $|x-x_0| > L$
(Fig. \ref{fig:current}).  The magnetic field resulting from the
current lies in the y-z plane, with variation in the $\bm{\hat
x}$ direction.  In our models, the heating region is near the center
of the outer crust, with characteristic width much smaller than the
crust thickness to ensure negligible heating at the boundaries.  We note that this current model leads to large pressure gradients in the crust.  To form a stable current model, a complex geometry is required, such as that of \cite{braithwaite09}.  We use the simplified current model described here to demonstrate the thermo-resistive instability. 

\subsection{Input Physics}

For the electrical and thermal conductivities in the crust, we use the analytical expressions of \cite{potekhin} for electron-ion collision frequencies in Fe matter.  We ignore the effects of the magnetic field on the conductivities.  Small jumps in the conductivities occur at $T=T_{melt}$, which we smooth to avoid numerical problems.
The heat capacity of the crust has contributions from ions and relativistic electrons.  The specific heat due to ions for solid matter ($\gamma > 150$) is given by \cite{vanriper},
  
\begin{equation}
   c_v^{ion} = n_i \, \frac{3}{2} k_b \left [ 1 + \frac{\log(\gamma)}{\log(150)} \right ] \,\,\,\rm{erg} \,\rm{cm}^{-3} \,\rm{K}^{-1}, 
\end{equation}
where 
\begin{equation}
\gamma \approx \frac{22.75 Z^2}{T_6} \left ( \frac{\rho_6}{A} \right
)^{1/3}
\end{equation}
is the Coulomb plasma parameter, $\rho_6$ is the density in units of $10^6 \rm{g} \, \rm{cm}^{-3}$, $T_6$ is the temperature in units of $10^6$ K, $Z$ is the ionic charge, and $A$ the atomic weight.  The contribution from relativistic, degenerate electrons to the specific heat is
\begin{equation}
  c_v^e = 5.4 \times 10^{19} \left(\frac{n_e}{n_0} \right )^{2/3} \,
  T_9 \,\,\, \rm{erg} \,\rm{cm}^{-3} \,\rm{K}^{-1},
\end{equation}
where $n_0=0.16 \,\rm{fm}^{-3}$ is the nuclear saturation density.  The specific heat of ions and electrons are shown in Fig. 4 at $10^8 $ K.

\begin{figure}
     \includegraphics[scale=0.5]{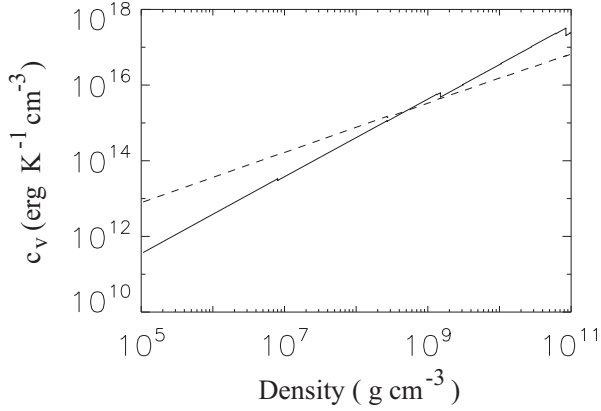}
     \caption{The specific heat at $10^8 \rm{K}$.  The solid line is ionic specific heat, and the dashed line is electronic specific heat.}
\end{figure}

In strongly-magnetized neutron star crusts, the neutrino luminosity is
dominated by neutrino synchroton emission \citep{aguilera}.  The rate
of emission is given by \cite{bezchastnov}
\begin{equation}
   Q_{\nu} = 10^{19} B_{15}^2 T_9^5 \,\,\, \rm{erg} \,\rm{cm}^{-3} \,\rm{s}^{-1},
\end{equation}
where $B_{15}$ is the magnetic field strength in units of $10^{15} $ G, and $T_9$ is the temperature in units of $10^9 $ K.  The ratio of neutrino emission $Q_{\nu}$ to ohmic heating $H$ is given by

\begin{equation}
    \frac{Q_\nu}{H} = 7 \times 10^{-5} \, B_{15}^2 \, T_8^5 \, \left ( \frac{\eta}{10^{-4} \, \rm{cm} \,\rm{s}} \right )^{-1} \, \left ( \frac{j}{10^{21} \,(\rm{erg} \,\rm{cm}^{-3} \,\rm{s}^{-2})^{1/2}} \right ) ^{-2}.
\end{equation} 
For the range of temperatures and magnetic fields we consider, the energy lost through neutrino emission is negligible compared to the ohmic heating rate and we neglect it in this analysis.  

\begin{figure}
     \includegraphics[scale=0.5]{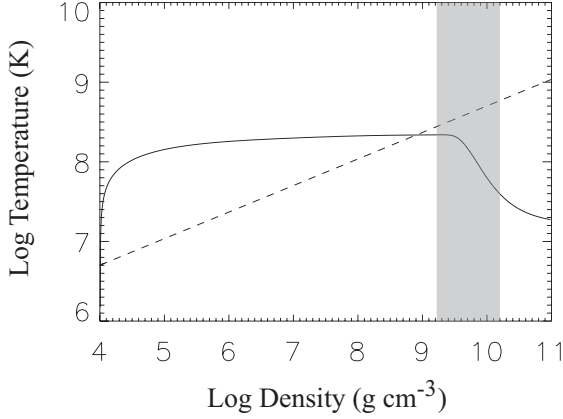} 
     \caption{A sample temperature profile, with $T_s = 10^6 $ K, $B_{max}= 5 \times 10^{15}$ G.  The dashed curve is the melting temperature of the lattice.  Shading indicates the heated region. This model indicates that a portion of the crust is molten, but the heated region is solid.}
\label{fig:tprofile}
\end{figure}

\begin{figure} 
     \includegraphics[scale=0.5]{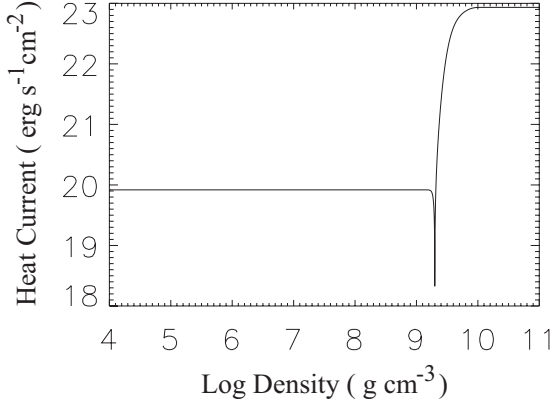}
     \caption{Absolute value of the heat current for a sample crust model.  The heat current flows towards the surface and the core from the heating peak.  Most of the flux is lost to the core. }
\label{fig:heatcurrent}
\end{figure}

Using these boundary conditions, we integrate eq. (\ref{eq:energy}) to the boundary of the outer crust.  An example of a typical equilibrium crust temperature profile is
plotted in Fig. \ref{fig:tprofile}, corresponding to the current
model shown in Fig. \ref{fig:current}.  The heat current for a
typical crust model is shown in Fig. \ref{fig:heatcurrent}.  
Most of the energy flux is into the star, in the direction of
increasing thermal conductivity. The energy is then lost to neutrinos
from the core.

\subsection{Stability Analysis}

We now examine the stability of the equilibrium state . We perform a
stability analysis of the outer crust using eq (\ref{eq:energy}), substituting $T(t)
= T_0 + \delta T e^{\gamma t}$, where $\delta T$ is the perturbation
mode and $\gamma$ is the growth (decay) rate.  For the range of
heating models we consider, instability growth rates are fast compared
to magnetic diffusion and we neglect evolution of the magnetic field
in our calculations.  Appendix A contains further justification of
this approximation.  Neglecting magnetic evolution and neutrino emission, the perturbed
energy balance equation is given by
\begin{equation} \label{eq:perturb}
   \frac{\partial^2 \delta T}{\partial x^2} = \frac{1}{\kappa} \left [ c_v \gamma  -  \frac{4 \pi}{c^2} \eta' j_0^2 -  \frac{\partial \kappa'}{\partial x} \frac{\partial T_0}{\partial x} - \kappa ' \frac{\partial^2 T_0}{\partial x^2} \right ] \delta T  -  \frac{1}{\kappa} \left [ \frac{\partial \kappa_0}{\partial x} + \kappa ' \frac{\partial T_0}{\partial x} \right ] \frac{\partial \delta T}{\partial x},    
\end{equation}
where primes indicate differentiation with respect to $T$ at fixed $x$.

\emph{Perturbation mode boundary conditions.}
To determine the perturbation mode gradient at the outer boundary, ($\rho=10^6 g \, cm^{-3}$), we integrate through the atmosphere for several values of $T_s$ to determine the dependence of the temperature gradient on the temperature.  Because there is no ohmic heating in the atmosphere section, the temperature at the outer boundary $T_{ob}$ and the temperature gradient ${dT_{ob}}/{dx}$ are uniquely defined for a given surface temperature.  Therefore, ${d T_{ob}}/{dx}$ is a function of $T_{ob}$:

\begin{equation}
   \frac{dT_{ob}}{dx} = f(T_{ob})
\end{equation}

Allowing perturbations to the temperature at the outer boundary for eq. (20) gives
\begin{equation}
    \frac{d (T_{ob} + \delta T_{ob})}{dx} = f(T_{ob} + \delta T_{ob}).
\end{equation}
Since the function $f(T_{ob})$ is well behaved, eq (21) to first order in $\delta T_{ob}$ is

\begin{equation}
    \frac{ d(T_{ob} + \delta T_{ob})}{dx} = f(T_{ob}) + f'(T_{ob}) \delta T_{ob},   
\end{equation}
where primes indicate differentiation with respect to $T_{ob}$.  Using eq. (20) we can solve for $f'(T_{ob})$,

\begin{equation}
   f'(T_{ob}) = \frac{d}{dT_{ob}} \left ( \frac{dT_{ob}}{dx} \right ).
\end{equation}
Keeping only perturbed terms of eq. (22), we arrive at the outer boundary condition,

\begin{equation}
   \frac{d \delta T_{ob}}{dx} = \frac{d}{dT_{ob}} \left (  \frac{dT_{ob}}{dx} \right ) \delta T_{ob}.
\end{equation}

We assume the inner crust/core of the neutron star to be an infinite heat reservoir.  Therefore, at the inner boundary we require that the temperature perturbation vanish, 

\begin{equation}
\delta T_{ib}= 0.
\end{equation}
We solve eq. (\ref{eq:perturb}) simultaneously for the temperature perturbation mode $\delta T(x)$ and the mode growth rate $\gamma$ for a given crust temperature profile and current distribution, given by eq. (9).  We satisfy the set of mixed boundary conditions through a shooting algorithm.   A sample temperature perturbation mode for the heating model presented previously is plotted in Fig. 7.  The instability growth rate $\gamma$ as a function of the maximum field in the crust due to crustal currents is plotted in Fig. 8.  

\begin{figure}
     \includegraphics[scale=0.5]{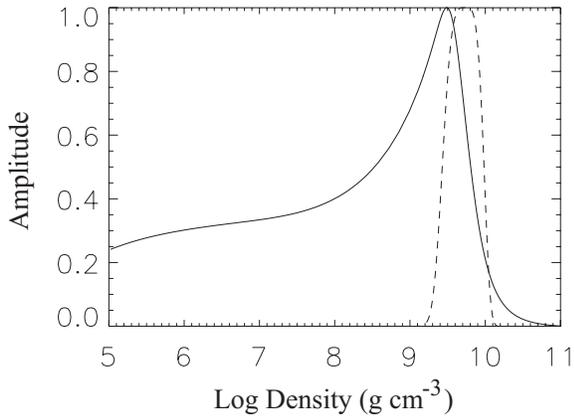} \caption{Sample unstable
     temperature perturbation mode, corresponding to the same crust
     model as Fig. 6. Dashed curve indicates the (normalized)
     electric current amplitude. Units are arbitrary.}
\end{figure}

\begin{figure}
     \includegraphics[scale=0.5]{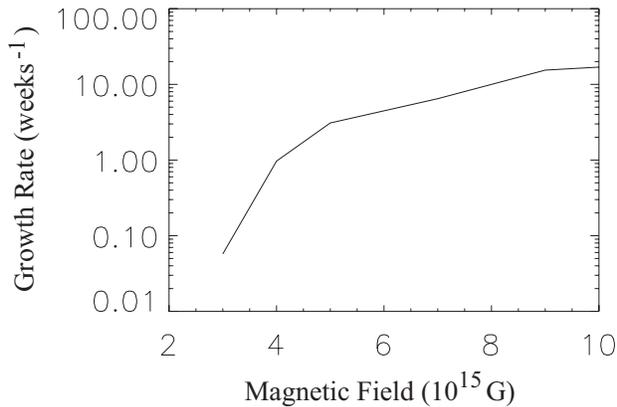}
     \caption{Instability growth rate $\gamma$ vs maximum field in the crust, for heating peaked at $\rho=3 \times 10^9 \rm{g} \, \rm{cm}^{-3}$.  The minimum field required for instability is $B=3 \times 10^{15}$ G.}
\end{figure}

To see the scalings of $\gamma$ with the parameters of the
problem, it is useful to perform a local plane wave analysis to obtain
an approximate analytical expression. Solving eq.
(\ref{eq:perturb}), assuming that $\kappa$ has only weak dependence on
$T$ and $x$ and keeping only the dominant terms (determined by numerical experiment),
we obtain
\begin{equation}
     \frac{\partial^2 \delta T}{\partial x^2} = \frac{1}{\kappa} \left [ c_v \gamma  -  \frac{4 \pi}{c^2} \eta' j_0^2 \right ] \delta T.
\end{equation}
Substituting plane-wave solutions for $\delta T$ with $k=1/L$ gives an approximate expression for the mode growth rate,

\begin{equation} \label{eq:gamma}
  \gamma \sim \frac{1}{c_v} \left [ \frac{4 \pi}{c^2} \eta_p' j_p^2 - \frac{\kappa_p}{L^2} \right ],
\end{equation}
where all parameters are evaluated at the heating peak.  Eq. (\ref{eq:gamma}) shows the competition between the ohmic heating (the first term), and cooling through thermal diffusion (the second term).  Changing the surface temperature affects the growth rate by changing the crust temperature at the heating location.  If the crust temperature is much greater than $T_{melt}$, the temperature sensitivity of the resistivity $\eta$ becomes negligible and the heating feedback effect is lost, thereby stabilizing the system.  Increasing the current amplitude gives more thermal energy to drive the instability.  The growth rate is somewhat insensitive to the choice of $L$.  However, the relationship between the magnetic field and the current (eq. (12)) indicates that for a fixed current amplitude $j_0$, larger values of $L$ correspond to larger magnetic fields.  To obtain crust models with realistic magnetic field amplitudes, the electric current must be concentrated in a relatively small region of the crust.  Finally, the instability growth rate is highly dependent on the heating location $x_0$ due to the spatial variation of both the thermal conductivity and the electrical resistivity.  A fixed current will produce the most heat in regions of high resistivity.  Heat produced in regions of low thermal conductivity is most likely to lead to instability, as the heat is not efficiently carried away.  For these reasons, heating near the surface of the star is most unstable.  

We calculate the instability growth rate for a wide range of heating models - characteristic heating widths range from $ 5 \, \rm{m}  < L  < 50 \, \rm{m}$, and heating peak locations $10^8 \, \rm{g} \, \rm{cm}^{-3} <  \rho_0 < 10^{10} \, \rm{g} \, \rm{cm}^{-3}$.  In this range, the current amplitude is virtually zero at the inner and outer boundaries.  We consider current amplitudes corresponding to maximum magnetic fields from $10^{15} $ G to $10^{16} $ G.  The steady state surface temperature $T_s$ is not the same for each heating model.  We calculate crust temperature profiles with $10^6$ K  $< T_s$  $< 10^7$ K, seeking models for which the temperature in the heating region is less than or equal to the melting temperature, and the temperature at the inner boundary is $10^7$ K  $< T_{ib}$  $< 10^8$ K.  Maintaining a crust temperature below $T_{melt}$ ensures that the resistivity feedback is operating at the heating location, and  $10^7$ K  $< T_{ib}$  $< 10^8$ K ensures that the core temperature is in the appropriate range for a magnetar with a characteristic age of $10^4$ yr \citep{aguilera}.  These conditions lead to a narrow range of values for the surface temperature.  Fig. 9 illustrates the region of instability for given values of the maximum field, $B_{max}$, and the core temperature of the neutron star for fixed heating width and location.  Subject to the conditions mentioned above, our results show that a minimum crustal field of $B = 3 \times 10^{15} $ G is required to destabilize the crust.   Instability growth rates for several models are plotted in Fig. 10.  To find an approximate expression for the critical field required to give instability ($\gamma > 0$), we substitute $j \sim c B_0/4 \pi L$ in eq (\ref{eq:gamma}) and solve for the critical field,

\begin{equation}
  B_{\rm{crit}} = 2.5 \times 10^{15} \left( \frac{\kappa}{2 \times 10^{18} \, \rm{erg} \,\rm{s}^{-1} \, \rm{cm}^{-1} \, \rm{K}^{-1}} \right ) \, \left (\frac{\eta'}{4 \times 10^{-12} \,\rm{cm}^2 \,\rm{s}^{-1}} \right )^{-1} \,\,\, \rm{G}.
\end{equation}
This estimate agrees well with our numerical solutions.

\begin{figure}
     \includegraphics[scale=0.5]{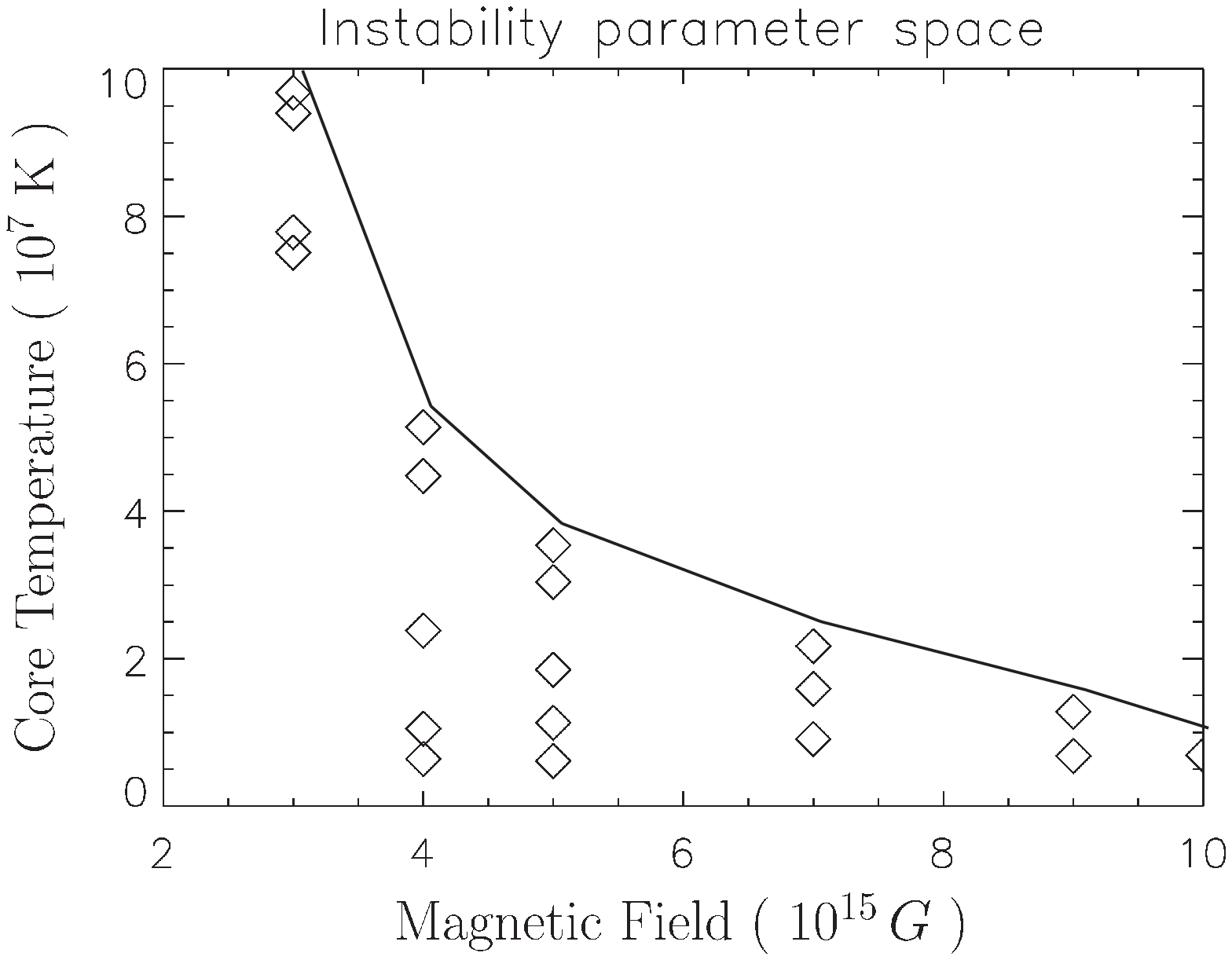}
     \caption{Diamonds represent values of the crust magnetic field and the neutron star core temperature for which unstable modes are found.  The solid line indicates the approximate boundary between unstable and stable parameter space.  All models used a heating width of 20 m, with heating location $\rho_0 = 3 \times 10^9 \rm{g} \rm{cm}^{-3}$. }
\end{figure}

\begin{figure}
     \includegraphics[scale=.52]{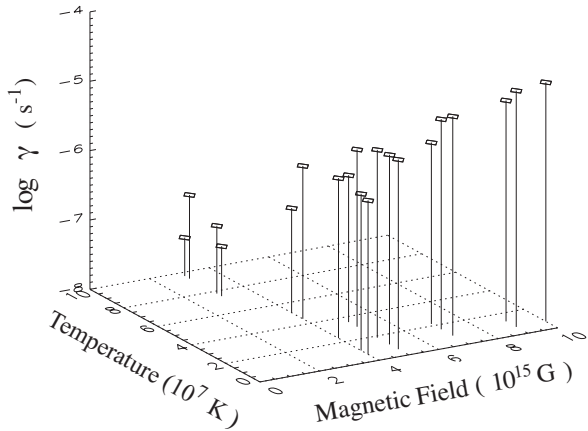}
     \caption{The instability growth rate as a function of the neutron star core temperature and the crust magnetic field.   All models used a heating width of 20 m, with heating location $\rho_0 = 3 \times 10^9 \rm{g} \rm{cm}^{-3}$ .}
\end{figure}

\section{Discussion and Conclusions}

Large crustal currents associated with the magnetic field of magnetars
may lead to a thermo-resistive instability in the crust.
Calculations of the instability growth time for a model neutron star crust give
typical growth times of weeks to months.  These timescales are short
compared to the ohmic diffusion timescale of the magnetic field. 

We conclude that the instability identified in this paper may operate
in neutron star crusts for a wide range of physical parameters relevant to
magnetars.  Heating may be located anywhere in the outer crust, while
magnetic length scales need only be comparable to the crust thickness
or smaller.  Instability occurs for crust temperatures
$T_{\rm{crust}} \sim 5 \times 10^8 $ K or lower, characteristic of
relatively young magnetars, $\tau_{\rm{age}} \sim 10^4 $ yrs
\citep{aguilera}.  This result coincides with the inferred age of
magnetar candidates associated with supernova remnants (see
\citet{mereghetti} for a review).  Additionally, we find that only
heating models that produce large magnetic fields ($B > 10^{15} $ G)
will produce instability, so this instability is specific to magnetars.  The characteristic temperatures and magnetic fields at which the thermo-resistive instability occurs suggest and intriguing connection between this instability and magnetars. We note that our simplified treatment of the current sheet is likely to overestimate the critical field required for instability by a factor of order unity. A stable current sheet necessarily has a more complicated structure than assumed here. The components of the current that we have neglected will lead to further heating. 

While we restrict our models to heating in the outer crust, 
instability in the inner crust could arise in the same way. 
Heat and charge transport mechanisms are no different, and we expect
the scaling of eq. (\ref{eq:gamma}) for the growth rate to apply to
inner crust heating.  However, because of the larger thermal
conductivity in the inner crust, deeper crustal currents would have to
be larger to produce similar instability growth rates to those
calculated here.  As the heating is moved to deeper layers, the minimum magnetic field required for instability becomes greater than $10^{16} $ G.  

The magnetic energy in the crust for our higher heating models is sufficient to power even the largest giant flares.  The magnetic energy contained in the magnetic field in the crust is given by
\begin{equation}
    E_B = 4 \pi \, \int  r^2 \frac{B(r)^2}{8 \pi} \, dr,
\end{equation}  
where we integrate from the inner boundary to the stellar
surface.  We calculate the maximum field in
the crust, the rate of energy deposition due to crustal currents, and
the magnetic energy in the outer crust for several heating models (Table 1). The energy released during the largest giant flare to date was approximately
$10^{46} $ erg. 

\begin{table}
\centering
\begin{tabular}{|c|c|c|c|}
\hline
 Model & B (G) & $\dot Q (\rm{erg} \rm{s}^{-1})$ & $E_B $ (erg) \\
\hline
1 & $10^{15}$ & $2 \times 10^{35}$ & $5.7 \times 10^{45}$ \\
2 & $5 \times 10^{15}$ & $10^{36}$ & $1.4 \times 10^{47}$ \\
3 & $10^{16}$ & $2 \times 10^{36}$ & $5.6 \times 10^{47}$ \\
\hline
\end{tabular}
\caption{Magnetic field, heat deposition rate and magnetic energy in the crust for 3 heating models.}  
\end{table}

Future work is required to determine the nonlinear evolution of the
magnetic field once an instability is triggered.  Solving the coupled
energy conservation and magnetic induction equations as a function of
time will give insight into this problem.  Our solutions for the steady state temperature indicate that a portion of the crust is molten.  As the instability grows,
higher density regions of the crust may melt, reducing the maximum magnetic
stress that could be supported by the crust.  Simulations of the magneto-thermal evolution could provide a link between the instability we have identified and glitch behavior in magnetars.

\vspace{40mm}
\Large
Appendix A - Magnetic Induction
\normalsize
\vspace{5mm}

In this Appendix we perform a local, plane-wave analysis of eqs. (19) and (20).  The perturbed energy conservation equation including magnetic induction is given by

\begin{equation}
   \kappa \frac{\partial^2 \delta T}{\partial x^2} = \left [ C_v \gamma - \frac{4 \pi}{c^2} \eta' j_0^2 - \frac{\partial \kappa'}{\partial x} \frac{\partial T_0}{\partial x} - \kappa ' \frac{\partial^2 T_0}{\partial x^2} \right ] \delta T - \left [ \frac{\partial \kappa_0}{\partial x} + \kappa ' \frac{\partial T_0}{\partial x} \right ] \frac{\partial \delta T}{\partial x} -  \left [ \frac{8 \pi}{c^2} \, \eta \, j_0 \right ] \delta j.
\end{equation}

The magnetic induction equation (eq. (2)) can be written in terms of the current $\bm{ j}$,

\begin{equation}
    \frac{\partial \bm{ j}}{\partial t} = - \bm{ \nabla} \times ( \bm{ \nabla} \times \eta \bm{ j}).
\end{equation}

Upon linearizing the induction equation, we have

\begin{eqnarray}
   \eta_0 \frac{\partial^2 \delta j}{\partial x^2} &=& \left [ \gamma - \frac{\partial^2 \eta_0}{\partial x^2} \right ] \delta j  -  2 \frac{\partial \eta_0}{\partial x} \frac{\partial \delta j}{\partial x} 
  - \left [2 \frac{\partial \eta '}{\partial x} \frac{\partial j_0}{\partial x} + \eta ' \frac{\partial^2 j_0}{\partial x^2} + \frac{\partial^2 \eta ' }{\partial x^2} j_0 \right ] \delta T \nonumber \\
 & &-  \left [ 2\frac{\partial \eta '}{\partial x} j_0 + 2 \eta ' \frac{\partial j}{\partial x} \right ] \frac{\partial \delta T}{\partial x} - \eta ' j_0 \frac{\partial^2 \delta T}{\partial x^2}. 
\end{eqnarray}

Magnetic induction can be neglected if the induction term is small compared to the ohmic heating term in eq. (31), integrated over the heating region:

\begin{equation}
   \int_V \frac{4 \pi}{c^2} \eta ' j_0^2 \gg \int_V \frac{8 \pi}{c^2}\, \eta_0 \,j_0 \,f(x),
\end{equation}
where $f(x) \equiv \delta j / \delta T$.  We find an approximate expression for $f(z)$ using plane wave solutions for $\delta T$ and $\delta j$ in eq. (33) and solving for $\delta j$:

\begin{equation}
   \delta j = \left [  {k^2 \eta_0 + \gamma - \frac{d^2 \eta_0}{dx^2} - 2 i k \frac{\eta_0}{dx} } \right ]^{-1}\left [{2 \frac{d \eta '}{dx}\frac{d j_0}{dx} + \eta' \frac{d^2 j_0}{dx^2} + \frac{d^2 \eta'}{d x^2}j_0 + 2 i k \frac{d \eta'}{dx} j_0 + 2 i k \eta' \frac{d j_0}{dx} - \eta' j_0 k^2 }\right ] \delta T.
\end{equation}

We take for the wavenumber $k=1/L$, the characteristic width of the heating region.  We approximate the induction term by evaluating the variables at the heating peak, 

\begin{equation}
   \int_V \frac{8 \pi}{c^2} \, \eta_0 \, j_0 \, f(x) \simeq \frac{8 \pi}{c^2} \, L \,\eta_p \,j_{max}\, f(x),
\end{equation}
where $L$ is the heating length scale, $\eta_p$ is the resistivity at the heating peak, $j_{max}$ is the maximum current, and $f(x)$ is evaluated at the heating peak in the plane wave approximation.  We approximate the heating term in the same way,

\begin{equation}
   \int_V \frac{4 \pi}{c^2} \eta ' j_0^2 \simeq \frac{4 \pi}{c^2} L_h \eta_p' j_{max}^2,
\end{equation}
where $\eta_p'$ is $\frac{\partial \eta}{\partial T}$ evaluated at the peak.  Using (18) and (19), eq. (16) becomes

\begin{equation}
     \frac{4 \pi}{c^2} L_h \eta_p' j_{max}^2 \gg  \frac{8 \pi}{c^2} \, L_h \,\eta_p \,j_{max}\, f(x).
\end{equation}

For the range of magnetic field models we consider, the heating term is much larger than the induction term.  We plot the ratio of the heating term to induction term in Fig. 11 for several models.  We see that magnetic induction is negligible for the large fields of interest.

\begin{figure}
     \includegraphics[scale=.50]{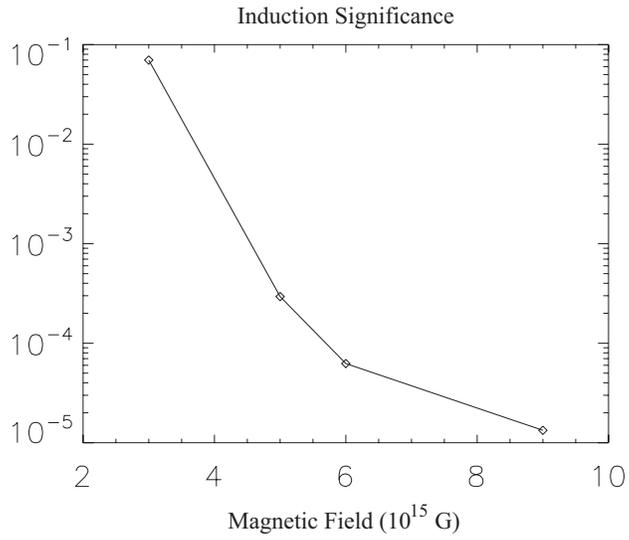}
     \caption{Ratio of induction term to heating term in the energy balance equation.  Induction can be neglected for small values of this ratio. }
\end{figure}

As a second test, we evaluate the ohmic decay time of the magnetic field by dimensional analysis.  Dimensionally, the ohmic decay time is

\begin{equation}
   \tau_d \sim \frac{L^2}{4 \pi \eta},
\end{equation}

where $\eta$ is the resistivity at the heating peak, and $L$ is a characteristic length scale.  Ohmic diffusion can be neglected if the ohmic decay timescale is much longer than the instability growth timescale, $\tau_d \gg \tau_g$.  Fig. 12 shows the ratio $\tau_d / \tau_g$ for several heating models.
\begin{figure}
     \includegraphics[scale=0.5]{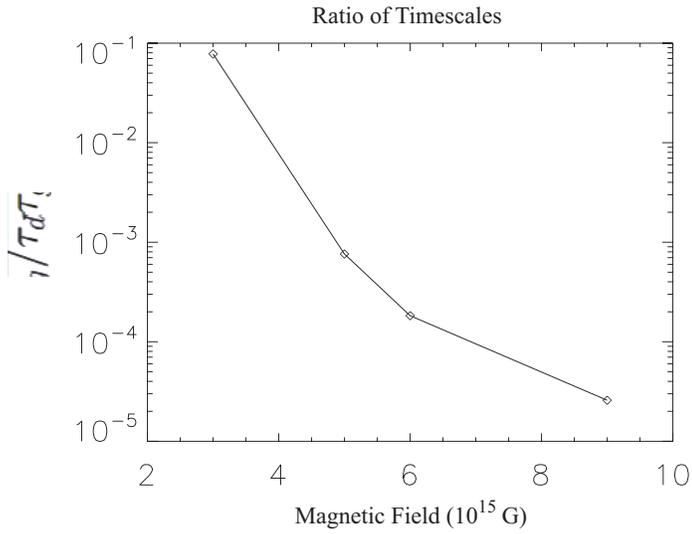}
     \caption{Ratio of instability growth timescale to ohmic decay timescale for several heating models.  Magnetic induction can be neglected for $\tau_g / \tau_d \ll 1$. }
\end{figure}

\bibliographystyle{apj}
\bibliography{myrefs2}	
\end{document}